\documentstyle[11pt,epsfig]{article}
\date{}
\title{\bf Coulomb's law in maximally symmetric spaces}
\author{ B. Vakili$^{1,}$\thanks{email: b-vakili@iauc.ac.ir}
\quad and \quad M. A. Gorji$^{2,}$\thanks{email:
m.gorji@mail.sbu.ac.ir }
\\\\{\small $^1${\it Department of Physics, Azad University in Chalous,
P. O. Box 46615-397, Chalous, Iran}}
\\{\small $^2${\it Department of Physics, Shahid Beheshti University, G. C. Evin,
Tehran 19839, Iran}}}

\begin{document}
\maketitle
\begin{abstract}
We study the modifications to the Coulomb's law when the
background geometry is a $n$-dimensional maximally symmetric
space, by using of the $n$-dimensional version of the Gauss'
theorem. It is shown that some extra terms are added to the usual
expression of the Coulomb electric field due to the curvature of
the background space.  Also, we consider the problem of existence
of magnetic monopoles in such spaces and present analytical
expressions for the corresponding magnetic fields and vector
potentials. We show that the quantization rule of the magnetic
charges (if they exist) would be applicable to our study as
well.\vspace{5 mm}\newline Keywords: Coulomb's law; maximally
symmetric space
\end{abstract}

\section{Introduction}
Almost all of the physical laws presented in the elementary text
books are formulated in a flat Euclidean space. Then, one of the
interesting questions which can be arise is that how these laws
may be modified if the corresponding physical systems live in a
curved space. Maybe the simplest example which can show the
effects of the curvature on the behavior of a physical system is
the issue of the parallel displacement of a vector in a curved
space. As is well-known in a flat space two vectors with the same
magnitude which are located in different points are same if they
are parallel. This means that in such a space we can parallel
transport a vector to everywhere without any change to its
magnitude and direction. On the other hand, parallel transporting
a vector from one point to another in a non-flat space yields a
result that depends on the path taken and specially when the path
is a closed loop the vector finally may not coincide with itself.
This effect is due to the curvature and therefore the dynamical
laws of a point particle (which are the relations between its
dynamical vectors) moving in a curved background may show some new
aspects which are absent in a flat space, see for example
\cite{Ga} and the references therein.

Here, we would like to look at the electrostatic force law between
charged particles. As we know, this interaction is governed by the
Coulomb's law according to which the force between point charges
obeys an inverse square rule. Now, the question is that how this
law may be corrected if we consider it in a curved space. The
first attempts to answer this question return to long ago in the
works of E. Fermi \cite{Fer}, where the electric field of a point
charge held at rest within a weak gravitational field has been
discussed. These efforts were further developed by others via
considering the electric phenomena in gravitational fields
\cite{Wh}.

In this letter we are going to obtain the modified version of the
Coulomb's law in a $n$-dimensional maximally symmetric curved
space. For this purpose we shall begin with the $n$-dimensional
version of the Gauss' theorem and see how the curvature effects
show themselves in the resulting expression for the electric field
and its corresponding scalar potential. We show that in the
vicinity of charges where the effect of curvature is negligible,
we recover usual Coulomb's law. Also, we consider the problem of
existence of the magnetic monopoles in our formalism. As is
well-known from the Dirac theory about these objects \cite{Dir},
an important features of them is that their existence would offer
an quantization rule for the magnitude of electric charges. We
shall see that this rule will be preserved when the background
geometry is a maximally symmetric curved space.
\section{Geometrical set-up}
In this section we briefly review the concept of the maximally
symmetric spaces used in the present work which originally
appeared in \cite{1}. As is well-known, the symmetry
transformations (isometries) of a given manifold, i.e., the
diffeomorphisms that leave the metric tensor invariant, are
characterized by the existence of the so-called Killing vector
fields and the number of such isometries are the number of
independent Killing vectors fields. Standard calculations in
differential geometry show that there can be at most
$\frac{1}{2}n(n+1)$ different independent Killing vector fields in
a space of dimension $n$. As an example, it is easy to show that
the Minkowski space admits $10$ Killing vectors, which is the
maximum number possible for a $4$-dimensional space-time. Also, in
a cosmological situation, the standard Friedmann-Robertson-Walker
models would be maximally symmetric in the sense that they would
admit all the $\frac{1}{2}n(n+1)$ different Killing vector fields.
In general, any $n$-dimensional maximally symmetric manifold
should be in the form of $S^n$ (the $n$-dimensional sphere), $P^n$
(the $n$-dimensional projective space), $E^n$ (the $n$-dimensional
Euclidean space) or $H^n$ (the $n$-dimensional hyperbolic space)
\cite{2}. It is worth to note that all of these spaces are
homogeneous and isotropic about all points. This is indeed a
generic property of the maximally symmetric spaces and one can
show that a given space is maximally symmetric if and only if it
is  homogeneous and isotropic. The study of the curvature
properties of such spaces shows that their Riemann-Christoffel
tensor takes the form
$\textit{R}_{ijkl}=K(g_{il}g_{jk}-g_{ik}g_{jl})$, where $K$ is a
constant called the curvature index. Spaces with this curvature
tensor are called spaces of constant curvature and their
geometries are quite different depending on whether the curvature
index is positive, negative or zero.

The construction of a $n$-dimensional maximally symmetric space
may be made by embedding it in a larger $(n+1)$-dimensional flat
space. In other word, a maximally symmetric space can be viewed
upon as a hypersurface in a flat $(n+1)$-dimensional space. To do
this, consider a flat $(n+1)$-dimensional space with line element
\begin{eqnarray}\label{A}
ds^2=g_{\mu\nu}dX^{\mu}dX^{\nu}, \qquad\qquad\qquad \mu,\nu =
1,2,....,n+1.
\end{eqnarray}
By introducing the constant $n\times n$ matrix $c_{ij}$ and also a
constant number $K$, the above line element may be rewrite as
follows
\begin{eqnarray}\label{B}
ds^2=c_{ij}dx^{i}dx^{j}+K^{-1}dw^2,    \qquad\qquad\qquad i,j =
1,2,....,n,
\end{eqnarray}
in which we used the symbol $w$ for the coordinate $X^{n+1}$. Now,
let us look at our desired maximally symmetric space and define it
as the following hypersurface
\begin{equation}\label{C}
Kc_{ij}x^{i}x^{j}+w^2=1.
\end{equation}
Depending on the sign of $K$ this hypersurface represents the
surface of a sphere (if $K>0$) or a pseudosphere (if $K<0$). Upon
substitution (\ref{C}) into relation (\ref{B}) we obtain

\begin{eqnarray}\label{D}
ds^2=c_{ij}dx^{i}dx^{j}+\frac{K(c_{ij}dx^{i}dx^{j})^2}{(1-Kc_{mn}dx^{m}dx^{n})},
\end{eqnarray}
which means that the metric functions induced on the hypersurface
can then be written in terms of $c_{ij}$ and $K$ as

\begin{equation}\label{E}
g_{ij}=c_{ij}+\frac{K(c_{il}c_{jk}x^{l}x^{k})}{(1-Kc_{mn}dx^{m}dx^{n})}.
\end{equation}
Now, it is straightforward to compute the corresponding
Riemann-Christoffel tensor with result
\begin{equation}\label{F}
R_{ijkl}=K(g_{il}g_{jk}-g_{ik}g_{jl}),
\end{equation}
which has the desired form for that the underlying geometry to be
maximally symmetric. Note that contracting once yields
\begin{equation}\label{G}
R_{ij}=K(n-1)g_{ij},\end{equation}where $R_{ij}$ is the Ricci
tensor corresponds to the metric $g_{ij}$. We see that the Ricci
tensor is proportional to the metric. Contracting once more gives
us the following relation between the Ricci scalar $R$ and the
curvature index
\begin{equation}\label{H}
K=\frac{R}{n(n-1)}.
\end{equation}

Before going any further, let us consider the structure of a
$n$-dimensional maximally symmetric space in a special case in
which $c_{ij}=K^{-1}\delta_{ij}$ if $K\neq 0$ and
$c_{ij}=\delta_{ij}$ if $K=0$. For this choice the metric
(\ref{D}) takes the form
\begin{eqnarray}\label{I}
ds^2=|K|^{-1}\left[dx_{i}dx^{i}+\frac{K|K|^{-1}}{1-K|K|^{-1}(x_{i}x^{i})}(x_{i}dx^{i})^2\right],
\end{eqnarray}for $K\neq 0$ and
\begin{equation}\label{M}
ds^2=dx_idx^i, \end{equation}for $K=0$. To simplify the above form
of the metric, let us introduce a new set of variables

\begin{eqnarray}\label{J}
\left\{
\begin{array}{ll}
x_1=R\cos \theta_1,\\\\
x_2=R\sin \theta_1 \cos \theta_2,\\\\
\quad.\quad.\quad.\quad.\quad.\quad.\quad.\quad.\quad,\\\\
x_{n-1}=R\sin \theta_1 \sin \theta_2...\sin \theta_{n-2}\cos
\theta_{n-1},\\\\
x_n=R\sin \theta_1 \sin \theta_2...\sin \theta_{n-2}\sin
\theta_{n-1},
\end{array}
\right.
\end{eqnarray}

where $R^{2}=x_{i}x^{i}$. In terms of these new variables the line
element (\ref{I}) takes the form

\begin{eqnarray}\label{K}
ds^2=|K|^{-1}\left(\frac{dR^{2}}{1-\textit{k}R^{2}}+R^{2}d\Omega_{n-1}^2\right),
\end{eqnarray}
where $k=\frac{K}{|K|}$ and
\begin{eqnarray}\label{L}
d\Omega_{n-1}^2=d\theta_1^2+\sin^2 \theta_1 d\theta_2^2+...+\sin^2
\theta_1\sin^2\theta_2...\sin^2\theta_{n-2}d\theta_{n-1}^2.
\end{eqnarray}
Notice that for $K\neq 0$ the parameter $k$ takes the values $\pm
1$ and for $K=0$ we identify $k=0$. In the following we
investigate in more detail the geometry of the above space for
three cases $k=0, +1$ and $-1$.

$\bullet$ Case $k=0$:

In this case from (\ref{M}) we have

\begin{eqnarray}\label{N}
ds^2=dR^{2}+R^{2}d\Omega_{n-1}^2=dx_{1}^{2}+dx_{2}^{2}+......+dx_{n}^{2},
\end{eqnarray}
which represents a $n$-dimensional Euclidean space and is called
flat space.

$\bullet$ Case $k=+1$:

Taking $k=+1$ in (\ref{K}) yields

\begin{eqnarray}\label{O}
ds^2=\frac{dr'^{2}}{1-|K|r'^{2}}+r'^{2}d\Omega_{n-1}^2,
\end{eqnarray}in which we have defined $r'$ by
$R=r'|K|^{1/2}$. Since the coefficient of $dr'^2$ becomes singular
as $r'\rightarrow |K|^{-1/2}$, we introduce a new coordinate
$r=\int\frac{dr'}{\sqrt{1-|K|r'^{2}}}$ which transforms (\ref{O})
into the form

\begin{eqnarray}\label{P}
ds^2=dr^{2}+|K|^{-1}\sin^{2}\left(\sqrt{|K|}r\right)
d\Omega_{n-1}^2.
\end{eqnarray}
This space has a compact topology and we call it as a closed or
spherical space $S^n$. Now, it is clear that the
$(n-1)$-hypersurfaces $r=\mbox{cons.}$ are $(n-1)$-spheres
$S^{n-1}$ with line element
\begin{eqnarray}\label{Q}
d\sigma^2=|K|^{-1}\sin^{2}\left(\sqrt{|K|}r\right)
d\Omega_{n-1}^2.
\end{eqnarray}

$\bullet$ Case $k=-1$:

Following the same steps as in the previous case but now with the
variable  $r=\int\frac{dr'}{\sqrt{1+|K|r'^{2}}}$ we are led to the
metric of the $n$-dimensional hyperbolic space as

\begin{eqnarray}\label{R}
ds^2=dr^{2}+|K|^{-1}\sinh^{2}\left(\sqrt{|K|}r\right)
d\Omega_{n-1}^2,
\end{eqnarray}which has obviously an open topology. Again the
$(n-1)$-hypersurfaces  are specified by
\begin{eqnarray}\label{S}
d\sigma^2=|K|^{-1}\sinh^{2}\left(\sqrt{|K|}r\right)
d\Omega_{n-1}^2.
\end{eqnarray}

\section{Coulomb's law}
As is well-known the interaction between charged particles can be
described by the Coulomb's law according to which any two pair of
such particles exert a force on each other. This force is a
central force which is proportional to the magnitude of each
charge and varies inversely as the square of the mutual distance.
Furthermore, this is an attractive force if the charges are unlike
and the like charges repel each other. According to these
statements the force exerted by point charge $q$ located at ${\bf
r'}$ on charge $q_0$ located at ${\bf r}$ can be written as
\begin{equation}\label{T}
{\bf E}({\bf r})=\frac{{\bf F}}{q_0}=\frac{1}{4\pi
\epsilon_0}\frac{q}{r^2}{\bf \hat{r}},\end{equation}where ${\bf
\hat{r}}$ is the unit vector directed from the charge $q$ to the
second charge $q_0$ and $r=|{\bf r}-{\bf r'}|$. Here, ${\bf E}$ is
is the electric force per unit charge and usually is called the
electric field strength. Now, assume that a field is set up by a
system of point charges, then according to the superposition
principle the total electric field strength equals the vector sum
of the electric fields due to the individual charges. One of the
important features of an electrostatic field strength is the
equality to zero of its curl at every point of the field , that
is, $\nabla \times {\bf E}({\bf r})=0$. Therefore, the strength of
an electrostatic field can be represented as the gradient of a
scalar function $\phi({\bf r})$ : ${\bf E}({\bf r})=-\nabla
\phi({\bf r})$. The function $\phi({\bf r})$ is known as the
potential of an electrostatic field and for a point charge $q$ is
equal to $\phi ({\bf r})=\frac{1}{4\pi \epsilon_0}\frac{q}{r}$.

From the general courses of physics we are familiar with Gauss'
theorem which for a field in vacuum can be worded as follows: the
flux of the vector field ${\bf E}$ through a closed surface is
proportional to the algebraic sum of the charges confined within
the surface, that is
\begin{equation}\label{U}
\oint_S {\bf E}.d{\bf s}=\frac{q}{\epsilon_0}.\end{equation}

In electrostatics coulomb's law may be deduced from Gauss' theorem
and vice versa. In this section we are going to obtain a modified
version of the Coulomb's law when the background geometry is a
$n$-dimensional maximally symmetric space. To do this, we shall
use the $n$-dimensional version of the Gauss' theorem \cite{3}.
Suppose that the point charge $q$ is located at the origin of the
coordinate system in a $n$-dimensional maximally symmetric space,
then according to the  $n$-dimensional Gauss' theorem the flux
through the hypersurface $S_{n-1}$ is given by

\begin{eqnarray}\label{V}
\oint_{S_{n-1}} {\bf E}_n.d{\bf S}_{n-1}=\frac{q}{\epsilon_{n}},
\end{eqnarray}
where $\epsilon_{n}$ is the permittivity of the $n$-dimensional
free space and dimensionally may be related with its three
dimensional counterpart $\epsilon_0$ as
$[\epsilon_n]=[\epsilon_0]/[L^{n-3}]$. Since a maximally symmetric
space is homogeneous and isotrope we can assume that ${\bf
E}_n={\bf E}_n({\bf r})$. Now, using (\ref{V}) we have

\begin{eqnarray}\label{X}
E_{n}(r)=\frac{q}{\epsilon_{n}\oint d S_{n-1}}.
\end{eqnarray}
What remain is to compute the area of the hypersurface over which
we would like to take the above integral. Using the equality
\begin{eqnarray}\label{Y}
S_{n-1}(Every Radius)=(Radius)^{n-1} S_{n-1}(Radius=1),
\end{eqnarray}and

\begin{eqnarray}\label{Z}
S_{n-1}(Radius=1)= \int_0^\pi \sin^{n-2}\theta_1
d\theta_1\int_0^\pi \sin^{n-3}\theta_2 d\theta_2...\int_0^\pi \sin
\theta_{n-2}d\theta_{n-2}\int_0^{2\pi}d\theta_{n-1}
=\frac{2\pi^{\frac{n}{2}}}{\Gamma(\frac{n}{2})},
\end{eqnarray}
we obtain the following relations for the area of a
$(n-1)$-hypersurface

\begin{eqnarray}\label{AB}
S_{n-1}=\left\{
\begin{array}{ll}
\frac{2\pi^{\frac{n}{2}}}{\Gamma(\frac{n}{2})}\left[|K|^{-\frac{1}{2}}\sin(\sqrt{|K|}r)\right]^{n-1},\hspace{.5cm}
\mbox{spherical space},\\\\\\
\frac{2\pi^{\frac{n}{2}}}{\Gamma(\frac{n}{2})}\left[|K|^{-\frac{1}{2}}\sinh(\sqrt{|K|}r)\right]^{n-1},\hspace{.5cm}\mbox{hyperbolic space}.\\
\end{array}
\right.
\end{eqnarray}
With these results at hand, we can compute the electrostatic field
strength related to a point charge placed at the origin of a
$n$-dimensional maximally symmetric space as

\begin{eqnarray}\label{AD}
E_n(r)=\left\{
\begin{array}{ll}
\frac{q \Gamma(\frac{n}{2})} {2\pi^{\frac{n}{2}} \epsilon_{n}
\left[|K|^{-\frac{1}{2}}\sin(\sqrt{|K|}r)\right]^{n-1}},\hspace{.5cm}
\mbox{spherical space},\\\\\\
\frac{q\Gamma(\frac{n}{2})} {2\pi^{\frac{n}{2}} \epsilon_{n}
\left[|K|^{-\frac{1}{2}}\sinh(\sqrt{|K|}r)\right]^{n-1}},\hspace{.5cm}\mbox{hyperbolic space}.\\
\end{array}
\right.
\end{eqnarray}

Let us focus on expressions (\ref{AD}) which represents the
electric field in a maximally symmetric curved space. Their
expansions in terms of the $r$ powers results

\begin{eqnarray}\label{AF}
E_{n}(r)= \frac{\Gamma(\frac{n}{2})q} {2\pi^{\frac{n}{2}}
\epsilon_{n} r^{n-1}}\left[1\pm\frac{(n-1)|K|}{6}r^2+
\frac{(5n-3)(n-1)|K|^{2}}{360}r^4 + ... \right],
\end{eqnarray}where the positive and negative signs correspond to
the spherical and hyperbolic spaces respectively. The first term
on the right hand side, is the Coulomb's law in a flat
$n$-dimensional space (see \cite{4}) and the other terms show the
effects of the curvature. In three dimension the above relation
takes the  following more familiar form

\begin{eqnarray}\label{AG}
E_{3}(r)=\frac{q}{4 \pi \epsilon_{0} r^2}\pm \frac{|K|q}{12 \pi
\epsilon_{0} }+\frac{|K|^{2}qr^2}{60 \pi \epsilon_{0} }+...
\end{eqnarray}
This is the modified form of the Coulomb's law in
three-dimensional maximally symmetric spaces which its
$r=\mbox{cons}.$ hypersurfaces have the metric $d\Omega^{2}=
a^{2}( d\theta_1 ^{2}+\sin^{2}\theta_1 d\theta_2 ^{2} )$. Since
the curvature index for this hypersurface is $K=1/a^{2}$, we get

\begin{eqnarray}\label{AH}
E_{3}(r)=\frac{q}{4 \pi \epsilon_{0} r^2}\pm \frac{q}{12 \pi
\epsilon_{0} a^2}+\frac{q r^2}{60 \pi \epsilon_{0} a^4}+...
\end{eqnarray}
As we have mentioned above, the first term is nothing but the
usual Coulomb's law in flat space. The second term although
depends on the curvature, does not depend on the distance. It
seems that this term shows a global effect of the curvature.
Finally, each of the remaining terms denote the local effects of
the curvature on the Coulomb's law. Now, let us to deal with the
scalar potential for the electric field calculated above. From the
standard definition

\begin{equation}\label{AI}
\phi_{n}(r) =- \int {\bf E}_{n}.{\bf dr},
\end{equation}

and by using of the relations (\ref{AD}) we arrive at the solution

\begin{eqnarray}\label{AJ}
\phi_n(r)=\left\{
\begin{array}{ll}
\frac{\Gamma(\frac{n}{2})q} {2\pi^{\frac{n}{2}} \epsilon_{n}
|K|^{-\frac{n}{2}}}\cos(\sqrt{|K|}r)\quad
F_{1\hspace{-.6cm}2}\hspace{.4cm}\left(\frac{1}{2} , \frac{n}{2},
\frac{3}{2}; \cos^{2} (\sqrt{|K|}r)\right) + C_{+},\hspace{.5cm}
\mbox{spherical space},\\\\\\
\frac{(-1)^{\frac{n}{2}}\Gamma(\frac{n}{2})q}{2\pi^{\frac{n}{2}}
\epsilon_{n} |K|^{-\frac{n}{2}}} \cosh(\sqrt{|K|}r)\quad
F_{1\hspace{-.6cm}2}\hspace{.4cm}\left(\frac{1}{2},\frac{n}{2},\frac{3}{2};
\cosh^{2} (\sqrt{|K|}r)\right) + C_{-},\hspace{.3cm}\mbox{hyperbolic space},\\
\end{array}
\right.
\end{eqnarray}
where $F_{1\hspace{-.5cm}2}\hspace{.4cm}(a,b,c;z)$ is
hypergeometric function and $C_{\pm}$ are some constants. In the
case of $n=3$ where $K=1/a^{2}$, the scalar potential for a three
dimensional maximally symmetric space takes the form
\begin{eqnarray}\label{AM}
\phi_3(r)=\left\{
\begin{array}{ll}
\frac{q}{4\pi \epsilon_0 a}\cot
\left(\frac{r}{a}\right)+C_{+},\hspace{.5cm}
\mbox{spherical space},\\\\
\frac{q}{4\pi \epsilon_0 a}\coth
\left(\frac{r}{a}\right)+C_{-},\hspace{.3cm}\mbox{hyperbolic space}.\\
\end{array}
\right.
\end{eqnarray}
As before the above relation can be expanded in the following way

\begin{eqnarray}\label{AN}
\phi_{3}(r)=C_{\pm}+\frac{q}{4 \pi \epsilon_{0}r}\mp\frac{q r}{12
\pi \epsilon_{0} a^{2}} - \frac{q r^{3}}{180 \pi \epsilon_{0}
a^{4} } + ...,
\end{eqnarray}
where the upper and lower signs correspond to the spherical and
hyperbolic spaces respectively. The discussions on the comparison
between different terms of this equation and their corresponding
effects are the same as previous, namely similar discussion based
on (\ref{AH}) would be applicable to this case as well.

\section{Magnetic monopole}
In this section we shall deal with the problem of magnetic
monopole in a curved maximally symmetric space. This issue is one
of the most interesting predictions of quantum mechanics which so
far had not been established experimentally. As is well-known from
elementary courses in electrodynamics, although there is a strong
symmetry between electric field ${\bf E}$ and magnetic field ${\bf
B}$ in Maxwell equations, but unlike the electric charges their
magnetic counterparts known as magnetic monopoles do not enter
these equations. Indeed, the source of all observable magnetic
fields in the nature is either moving electric charges or magnetic
dipoles and there is no evidence for the existence of magnetic
monopoles. This means that instead of the relation $\nabla . {\bf
B}=\rho_m$ which is the magnetic version of $\nabla . {\bf
E}=\rho_e$, in the usual way of writing Maxwell equations $\nabla
. {\bf B}$ is equal to zero. As a remark, we would like to
emphasize that quantum mechanics does not predict that the
magnetic monopole should exist but in a clear way implies that if
it exists the magnitude of the magnetic charges should obey a
quantization rule \cite{5}. Here, we are going to show the
validity of this statement in a curved maximally symmetric
background. So, let us suppose that there exist magnetic charge
density $\rho_m$. The Maxwell equation would then be $\nabla .{\bf
B}=\rho_m$. Therefore, if a point magnetic charge $e_m$ is placed
at the origin of a coordinate system, the magnetic counterpart of
the Gauss' law should be modified as

\begin{eqnarray}\label{AO}
\oint_{S_{n-1}}{\bf B}_{n}.d{\bf S}_{n-1}=e_{m}.
\end{eqnarray}
Following the same steps as in the previous section, this equation
lead us the magnetic field strength due the magnetic charge $e_m$
as

\begin{eqnarray}\label{AP}
B_n(r)=\left\{
\begin{array}{ll}
\frac{\Gamma(\frac{n}{2})e_m} {2\pi^{\frac{n}{2}}
\left[|K|^{-\frac{1}{2}}\sin(\sqrt{|K|}r)\right]^{n-1}},\hspace{.5cm}
\mbox{spherical space},\\\\
\frac{\Gamma(\frac{n}{2})e_m} {2\pi^{\frac{n}{2}}
\left[|K|^{-\frac{1}{2}}\sinh(\sqrt{|K|}r)\right]^{n-1}},\hspace{.3cm}\mbox{hyperbolic
space}.\end{array} \right.
\end{eqnarray}
Now, an essential question is that how can the magnetic fields
(\ref{AP}) be represented by a vector potential ${\bf A}$ such
that ${\bf B}=\nabla \times {\bf A}$. To deal with this question
let us restrict ourselves to the special case $n=3$ from now on
for which the relations (\ref{AP}) take the form

\begin{equation}\label{AT}
{\bf B}_{3}(r)= \frac{e_m }{4 \pi |K|^{-1}
\sin^{2}\left(\sqrt{|K|}r\right)} {\bf e}_r,
\end{equation}and

\begin{equation}\label{AU}
{\bf B}_{3}(r)= \frac{e_m }{4 \pi |K|^{-1}
\sinh^{2}\left(\sqrt{|K|}r\right)} {\bf e}_r,
\end{equation}where ${\bf e}_r$ is the local unit vector in
$r$-direction. On the other hand, bearing in the mind the general
relation
\begin{eqnarray}\label{AV}
\nabla \times {\bf A}=\frac{1}{h_{2} h_{3}}[
\partial_{2} (h_{3}A_{3}) - \partial_{3}(h_{2} A_{2})]{\bf e}_1 +
\frac{1}{h_{1}h_{3}} [ \partial_{3}(h_{1} A_{1}) -
\partial_{1} (h_{3} A_{3})]{\bf e}_2 + \frac{1}{h_{1}h_{2}} [
\partial_{1}(h_{2} A_{2}) - \partial_{2} (h_{1} A_{1})]{\bf e}_3,
\end{eqnarray}for the curl of a vector field in a curvilinear
metric space with scale factors $h_1$, $h_2$ and $h_3$, we can
write the following expressions for $\nabla \times {\bf A}$ in a
three dimensional maximally symmetric space \footnote {We have
used the notation
$ds^2=dr^2+|K|^{-1}\sin^2(\sqrt{|K|}r)\left(d\theta ^2+\sin^2
\theta d\varphi ^2\right)$ and
$ds^2=dr^2+|K|^{-1}\sinh^2(\sqrt{|K|}r)\left(d\theta ^2+\sin^2
\theta d\varphi ^2\right)$ for the metrics of spherical and
hyperbolic spaces respectively.}

\begin{eqnarray}\label{AX}
\nabla \times {\bf A}= \frac{1}{|K|^{- \frac{1}{2}}
\sin\left(\sqrt{|K|}r\right)\sin\theta}
\left[\partial_{\theta}(\sin\theta A_{\varphi}) - \partial_{\varphi} A_{\theta}\right] {\bf e}_{r} \qquad\qquad\qquad\qquad\qquad\qquad \nonumber\\
+\frac{1}{|K|^{- \frac{1}{2}} \sin\left(\sqrt{|K|}r\right)
\sin\theta} \left[
\partial_{\varphi} A_{r} -
\partial_{r} \left(|K|^{- \frac{1}{2}} \sin(\sqrt{|K|}r) \sin\theta A_{\varphi}\right) \right]{\bf e}_{\theta} \qquad\qquad  \nonumber\\
+\frac{1}{|K|^{- \frac{1}{2}} \sin\left(\sqrt{|K|}r\right)} \left[
\partial_{r}\left(|K|^{- \frac{1}{2}} \sin(\sqrt{|K|}r)
A_{\theta}\right)-\partial_{\theta} A_{\theta}\right] {\bf
e}_{\varphi},
\end{eqnarray}for spherical space and

\begin{eqnarray}\label{AY}
\nabla \times {\bf A}=\frac{1}{|K|^{- \frac{1}{2}}
\sinh\left(\sqrt{|K|}r\right) \sin\theta}
\left[\partial_{\theta}(\sin\theta A_{\varphi}) - \partial_{\varphi} A_{\theta}\right] {\bf e}_{r} \qquad \qquad\qquad\qquad\qquad\qquad\qquad\qquad \nonumber\\
+\frac{1}{|K|^{- \frac{1}{2}} \sinh\left(\sqrt{|K|}r\right)
\sin\theta}\left[
\partial_{\varphi} A_{r}-\partial_{r}\left(|K|^{-\frac{1}{2}}
\sinh(\sqrt{|K|}r)\sin\theta A_{\varphi}\right)\right]{\bf e}_{\theta}  \qquad\qquad\qquad  \nonumber\\
+\frac{1}{|K|^{- \frac{1}{2}} \sinh\left(\sqrt{|K|}r\right)}\left[
\partial_{r}\left(|K|^{- \frac{1}{2}} \sinh(\sqrt{|K|}r)
A_{\theta}\right)-\partial_{\theta} A_{\theta}\right] {\bf
e}_{\varphi},
\end{eqnarray}for hyperbolic space. Since the magnetic fields
(\ref{AT}) and (\ref{AU}) only have the $r$ component, we suppose
that their corresponding vector potentials only have the azimuthal
component and suggest the following ansatz

\begin{eqnarray}\label{AZ}
{\bf A}=\frac{e_m}{4 \pi |K|^{\frac{-1}{2}}
\sin\left(\sqrt{|K|}r\right)}\frac{1- \cos\theta}{\sin\theta}{\bf
e}_{\varphi},
\end{eqnarray}for spherical space and

\begin{eqnarray}\label{BA}
{\bf A}=\frac{e_m}{4 \pi |K|^{\frac{-1}{2}}
\sinh\left(\sqrt{|K|}r\right)} \frac{1-
\cos\theta}{\sin\theta}{\bf e}_{\varphi},
\end{eqnarray}for hyperbolic space. A problem related to the above
potentials is that they are singular at $\theta=\pi$. Following
\cite{5}, to pass this problem, instead of an expression for the
vector potential which is valid everywhere we may define two
vector potentials ${\bf A}^{(I)}$ and ${\bf A}^{(II)}$ each of
which is valid in one of the regions $\theta<\pi-\delta$ and
$\theta>\delta$. Therefor, we modify the equations (\ref{AZ}) and
(\ref{BA}) as
\begin{eqnarray}\label{BC}
{\bf A}=\left\{
\begin{array}{ll}
{\bf A}^{(I)}=\frac{e_m}{4\pi |K|^{-\frac{1}{2}}\sin\left(\sqrt{|K|}r\right)}\frac{1-\cos \theta}{\sin \theta}{\bf e}_{\varphi},\hspace{.5cm}
\theta <\pi-\delta\\\\
{\bf A}^{(II)}=- \frac{e_m}{4 \pi |K|^{-\frac{1}{2}}
\sin\left(\sqrt{|K|}r\right)} \frac{1+
\cos\theta}{\sin\theta} {\bf e}_{\varphi},\hspace{.5cm}\theta>\delta,\\
\end{array}
\right.
\end{eqnarray}for spherical space and

\begin{eqnarray}\label{BD}
{\bf A}=\left\{
\begin{array}{ll}
{\bf A}^{(I)}=\frac{e_m}{4\pi
|K|^{-\frac{1}{2}}\sinh\left(\sqrt{|K|}r\right)}\frac{1-\cos
\theta}{\sin \theta}{\bf e}_{\varphi},\hspace{.5cm}
\theta <\pi-\delta\\\\
{\bf A}^{(II)}=- \frac{e_m}{4 \pi |K|^{-\frac{1}{2}}
\sinh\left(\sqrt{|K|}r\right)} \frac{1+
\cos\theta}{\sin\theta} {\bf e}_{\varphi},\hspace{.5cm}\theta>\delta,\\
\end{array}
\right.
\end{eqnarray}for hyperbolic space. It is seen that the potentials
$ {\bf A}^{(I)}$ and ${\bf A}^{(II)}$ together yield a correct
expression for the magnetic field which is valid everywhere. On
the other hand, since in the overlap region
$\delta<\theta<\pi-\delta$, these two potentials result the same
magnetic field, a gauge transformation of the kind ${\bf
A}^{(II)}\rightarrow {\bf A}^{(I)}+\nabla \Lambda$ may connect
them in this region \cite{6}. Noting that
\begin{eqnarray}\label{BE}
{\bf A}^{(II)}-{\bf A}^{(I)}=\left\{
\begin{array}{ll}
-\frac{e_m}{4\pi
|K|^{-\frac{1}{2}}\sin\left(\sqrt{|K|}r\right)}\frac{2}{\sin
\theta}{\bf e}_{\varphi},\hspace{.5cm}\mbox{spherical space} \\\\
-\frac{e_m}{4\pi
|K|^{-\frac{1}{2}}\sinh\left(\sqrt{|K|}r\right)}\frac{2}{\sin
\theta}{\bf e}_{\varphi},\hspace{.4cm}\mbox{hyperbolic space},
\end{array}
\right.
\end{eqnarray}and

\begin{eqnarray}\label{BE}
\nabla \Lambda=\left\{
\begin{array}{ll}
(\partial_r \Lambda_r){\bf e}_r +
\frac{1}{|K|^{-\frac{1}{2}}\sin\left(\sqrt{|K|}r\right)}
(\partial_{\theta} \Lambda_{\theta}) {\bf e}_{\theta} +
\frac{1}{|K|^{- \frac{1}{2}}\sin\left(\sqrt{|K|}r\right)
\sin\theta}
(\partial_{\varphi} \Lambda_{\varphi}){\bf e}_{\varphi},\hspace{.5cm}\mbox{spherical space}, \\\\
(\partial_r \Lambda_r){\bf e}_r +
\frac{1}{|K|^{-\frac{1}{2}}\sinh\left(\sqrt{|K|}r\right)}
(\partial_{\theta} \Lambda_{\theta}) {\bf e}_{\theta} +
\frac{1}{|K|^{- \frac{1}{2}}\sinh\left(\sqrt{|K|}r\right)
\sin\theta} (\partial_{\varphi} \Lambda_{\varphi}){\bf
e}_{\varphi},\hspace{.2cm}\mbox{hyperbolic space},
\end{array}
\right.
\end{eqnarray}

a simple calculation shows that
\begin{equation}\label{BF}
\Lambda=-\frac{e_m}{2\pi}\varphi.\end{equation}Now, consider a
charged particle of charge $e$ (say an electron) moving in the
magnetic field (\ref{AT}) or (\ref{AU}). Since in the overlap
region we can use either ${\bf A}^{(I)}$ or ${\bf A}^{(II)}$ the
corresponding wave functions obey a phase transformation of the
kind

\begin{eqnarray}\label{BG}
\Psi^{(II)}=\exp\left(\frac{ie \Lambda}{\hbar c}\right) \Psi^{(I)}
\end{eqnarray}
where $\hbar$ and $c$ are the Planck constant and  speed of light
respectively. If we demand that $\Psi$ should be a single-valued
function of $\varphi$, then we are led to

\begin{eqnarray}\label{BH}
e_m=\frac{2 \hbar c}{e} n, \hspace{0.5cm} n = 0, \pm 1, \pm 2, ...
\end{eqnarray}
which shows the quantization rule of the magnetic charges in a
curved maximally symmetric geometry is valid as well.

\section{Conclusions}
In this letter we have studied the modifications to the Coulomb's
law when the electric charges are living in a $n$-dimensional
maximally symmetric curved space. We obtained the general form for
the metric of such spaces and showed that they may be divided into
three classes which are called flat, spherical and hyperbolic
spaces. Then, by using of the $n$-dimensional version of the
Gauss' theorem we found analytic expressions for the electric
field intensity and scalar potential of a point charge placed at
the origin of the coordinate system. Our analysis showed that the
corresponding modifications have their origin in the curvature of
the background space and the usual inverse square Coulomb field
can be recovered in the limit where this curvature tends to zero.
We then dealt with the problem of the magnetic monopole in these
spaces and through calculating the corresponding magnetic field
and vector potential we arrived at the same quantization rule for
a magnetic charge as in their usual Dirac theory.
 \end{document}